\documentstyle[12pt]{article}

\begin{document}
\title{\textbf{Magnetic field reversals and topological entropy in non-geodesic hyperbolic dynamos}} \maketitle
{\sl \textbf{L.C. Garcia de Andrade}\newline
Departamento de F\'{\i}sica
Te\'orica-IF\newline
Universidade do Estado do Rio de Janeiro\\[-3mm]
Rua S\~ao Francisco Xavier, 524\\[-3mm]
Rio de Janeiro, RJ, Brasil\\[-3mm]
\\[-3mm]
\vspace{0.01cm} \newline{\bf Abstract} \paragraph*{Earlier, Chicone, Latushkin and Montgomery-Smith [Comm Math Phys (1997)] have proved the existence of a fast dynamo operator, in compact two-dimensional manifold, as long as its Riemannian curvature be constant and negative. More recently Gallet and Petrelis [Phys Rev \textbf{E}, 80 (2009)] have investigated saddle-node bifurcation, in turbulent dynamos as modelling for magnetic field reversals. Since saddle nodes are created in hyperbolic flows, this provides us with physical motivation to investigate these reversals in a simple kinematic dynamo model obtained from a force-free non-geodesic steady flow in Lobachevsky plane. Magnetic vector potential grows in one direction and decays in the other under diffusion. Magnetic field differential 2-form is orthogonal to the plane. A restoring forcing dynamo in hyperbolic space is also given. Magnetic field reversals are obtained from this model. Topological entropies [Klapper and Young, Comm Math Phys (1995)] are also computed.}
\newpage
\section{Introduction}
 It is well-known \cite{1} that every compact Riemann surface $\cal{M}$ can be endowed with a metric of constant curvature. This curvature is positive, for a sphere, zero for a torus and negative for any surface of genus (handles numbers) $g\ge{2}$. Arnold and Khesin \cite{2} showed that geodesic flows on every Riemann surface whose curvature is negative, and constant provides us with an example of a fast kinematic dynamo \cite{2} with dissipation where the magnetic field differential closed (dB=0) 2-form $\textbf{B}=d\textbf{A}$, where $\textbf{A}=A_{i}dx^{i}$ (i,j=1,2,3), does not depend on velocity. This non-geodesic flow defines a dynamical system in $\textbf{R}^{3}$ endowed with exponential stretching of particles. In this paper it is shown that non-geodesic magnetic flows embedded in a special type of Riemannian manifold of constant negative curvature 2D manifold called Lobachevsky plane, provides an example of a slow dynamo, where dynamo action cannot be supported in the absence of diffusion. The self-induction magnetic flow is solved, in the background of this hyperbolic space. In this case it is shown that the magnetic vector potential grows in one direction and decays in the other, showing some sort of anisotropy. On the other hand, an important one of the most important problems in geodynamos is the magnetic field reversal of the Earth \cite{3}. In this paper a forcing kinematic dynamo is used as a model for magnetic field reversal in hyperbolic manifold of negative constant Riemannian curvature. Another sort of flat embedding, that of Moebius strip into the $3D$ real Euclidean space $\textbf{E}^{3}$, has been used by Shukurov et al \cite{4} as a dynamo flow in a torus liquid sodium device in the Perm dynamo experiment. Actually the advantage of using the cartesian coordinates of the Lobachevsky plane is to simplify the computation and the solution of the self-induction equation instead of using the radial and angular coordinates of the paraboloid or hyperbolic space of constant Riemann negative curvature. Here it is also shown that non-Moebius thin surface flow with ellipsoidal cross-section the magnetic field embedded in the $\textbf{R}^{3}$ laboratory, decays in a non-dynamo profile. Unfortunately in some of these beautiful mathematical models of physical systems one main difficult is the embedding problems of the 2D negative constant Riemann curvature surfaces in 3D Euclidean space. Recently accomplished by Klebanov and Maldacena \cite{5} applied the idea of flat embedding of the $1+1$-D spacetime into a 2D negative constant curvature Riemannian hyperbolic space. This suggests that a fast or at least slow dynamo can be embedded in cosmological spatial section of the Friedmann-Goedel metric \cite{6}. Also recently the author has shown \cite{7} that the AdS anti-de Sitter spacetime leads to a slow dynamo when flat embedding is used in cosmology. The paper is organized as follows: Section II presents the fast dynamo in Lobachevsky plane. In this same section the non-geodesic equation is computed and the non-geodesy of the dynamo flow is characterized by the presence of an external force on the LHS of the geodesic equation. Section III, presents for the forcing dynamos and computation of topological entropy bounds for both examples is given in section IV. Future prospects and conclusions are presented in section V.
\newpage
\section{Geodesic deviation of stretched particles in fast dynamo hyperbolic flow}
In this section, one shall be concerned with the embedding of Lobachevsky plane given by the metric
\begin{equation}
ds^{2}=y^{-2}[dx^{2}+dy^{2}]
\label{1}
\end{equation}
where ${\Lambda}^{2}=(w=x+iy;y>0)$ is the hyperbolic plane in its half-upper part. Here $\sqrt{-1}=i$ is the imaginary unit of the complex plane $\textbf{C}$. Previously Sunada \cite{8} has given an example of a magnetic uniform (non-dynamo) field solution
\begin{equation}
{B}_{0}=y^{-2}dx{\wedge}dy
\label{2}
\end{equation}
where dx and dy are differential forms, on the Lobachevsky plane. Here one shall be concerned with the issue of a non-uniform time dependent magnetic flow on the Lobachevsky plane giving rise to a force-free dynamo flow. General dynamo hyperbolic flows can be useful to build spatial sections of cosmological models, as recently pointed out by Klebanov and Maldacena \cite{5} and Marteens et al \cite{9}. To shown that Klebanov and Maldacena Anti-de Sitter metric
\begin{equation}
ds^{2}=-e^{-2z}d{x_{0}}^{2}+dz^{2}\label{3}
\end{equation}
can be transformed into the Lobachevsky plane one simply consider the following transformation
\begin{equation}
z=lny\label{4}
\end{equation}
\begin{equation}
x=ix_{0}\label{5}
\end{equation}
Note that the magnetic flow can be defined as
\begin{equation}
{\Phi}_{t}:S^{1}({\Lambda}^{2}){\rightarrow}S^{1}({\Lambda}^{2})
\label{6}
\end{equation}
where $S^{1}$ is the sphere bundle with the group $SL_{2}(\textbf{R})$. The geodesic deviation (or non-geodesic) equation is given by 
\begin{equation}
\frac{d^{2}J}{ds^{2}}+K(s)J=0
\label{7}
\end{equation}
Thus since this equation, according to Arnold et al \cite{2} represents dynamo stretching of particles in the flow in the Riemannian spaces of negative curvature, a simple computation of the Gaussian curvature in terms of the Riemann curvature tensor component $R_{1212}$
\begin{equation}
K=\frac{R_{1212}}{g}
\label{8}
\end{equation}
Since for the Lobachevsky plane
\begin{equation}
{R_{1212}}=-y^{-4}
\label{9}
\end{equation}
and the determinant of the metric $g=y^{-4}$ one obtains a Gaussian curvature $K=-1$ which yields a simple geodesic deviation solution
\begin{equation}
J(s)=J_{0}sinh(\sqrt{-K}s)=J_{0}sinh(s)
\label{10}
\end{equation}
which shows that as $s\rightarrow{\infty}$ this particles separation J on the hyperbolic flow "explodes" in the form of a chaotic dynamo. In the next section one shall find an explicitly solution for the stretched magnetic field lines in the hyperbolic flow and the respective external force to support the slow dynamo action.
\section{Non-geodesic force-free fast dynamo hyperbolic flow and magnetic field reversal}
As is well-know the simple presence of magnetic fields guarantees the non-geodesic motion of the particles of the flow. After reviewing the state of the art of the dynamo problem in hyperbolic flows in the  last section, now let us consider a simple new solution of the self-induction equation
\begin{equation}
\frac{{\partial}\textbf{A}}{{\partial}t}+{\eta}{\nabla}^{2}\textbf{A}=\textbf{U}\times{\nabla}
\textbf{A}\label{11}
\end{equation}
where $\textbf{A}$ is the magnetic vector potential whose magnetic field is given by
\begin{equation}
\textbf{B}={\nabla}\times\textbf{A}\label{12}
\end{equation}
Since the frame of reference in this case is Euclidean this equation reduces to the following component equation
\begin{equation}
\frac{{\partial}{A^{i}}}{{\partial}t}+{\eta}{\nabla}^{2}{A^{i}}=[U_{j}(A^{i,j}-A^{j,i})]
\label{13}
\end{equation}
where ${\eta}\approx{Rm^{-1}}$ is the diffusion related with the magnetic Reynolds number $Rm$. Note that the Laplacian term is given in covariant coordinates by
\begin{equation}
{\nabla}^{2}A^{i}={\eta}{A^{i;k}}_{;k}\label{14}
\end{equation}
where the Riemannian covariant derivative Laplacian ${A^{i;k}}_{;k}$ is given by
\begin{equation}
{A^{i;k}}_{;k}=[g^{jk}({A^{i}}_{,k}+{{\Gamma}^{i}}_{lk}A^{l})]_{,j}+g^{jl}{{\Gamma}^{i}}_{
kj}({A^{k}}_{,l}+{{\Gamma}^{k}}_{lm}A^{m})+g^{kl}{{\Gamma}^{j}}_{
kj}({A^{i}}_{,l}+{{\Gamma}^{i}}_{lm}A^{m})\label{15}
\end{equation}
one also considers here that the Coulomb gauge
\begin{equation}
{\nabla}.\textbf{A}=\frac{{\phi}}{\eta}=0\label{16}
\end{equation}
since the electric field potential ${\phi}$ is assumed to vanish. Expansion of the equation in terms of the Riemann-Christoffel symbols ${\Gamma}_{ijk}$ above, and assumption that $A^{z}$ vanishes, reduces the self-induction equation to
\begin{equation}
\frac{{\partial}{A^{x}}}{{\partial}t}=[U_{y}(A^{x,y}-A^{y,x})]-{\eta}{\lambda}^{2}A^{x}
\label{17}
\end{equation}
and
\begin{equation}
\frac{{\partial}{A^{y}}}{{\partial}t}=[U_{x}(A^{y,x}-A^{x,y})]-{\eta}{\lambda}^{2}A^{y}
\label{18}
\end{equation}
To simplify the above solenoidal divergence-free equation
\begin{equation}
{{\partial}_{x}}(\sqrt{g}A^{x})+{\partial}_{y}(\sqrt{g}A^{y})=0
\label{19}
\end{equation}
one considers $A^{x}=A^{x}(y,t)$, which yields
\begin{equation}
{\partial}_{y}(\sqrt{g}A^{y})=0
\label{20}
\end{equation}
and the solution for $A^{y}$ is
\begin{equation}
A^{y}=A_{0}(t)y^{2}
\label{21}
\end{equation}
where a velocity flow as $v={V}^{y}\frac{{\partial}}{{\partial}y}$, has been used. Substitution of these solutions into the self-induction equation, (\ref{18}) one obtains
\begin{equation}
{A^{y}(y,t)}=A^{0}y^{2}e^{-{\eta}{\lambda}^{2}t}
\label{22}
\end{equation}
and
\begin{equation}
{A^{x}(y,t)}=A^{0}e^{{\gamma}t+Ky^{-2}}
\label{23}
\end{equation}
where ${\gamma}(\eta)=V_{0}K-{\lambda}^{2}{\eta}$, which shows that the dynamo is fast since $lim_{{\eta}\rightarrow{0}}{\gamma}(\eta)\ge{0}$. Here the magnetic field two-form \cite{2} is given by
\begin{equation}
B=B_{z}dx{\wedge}dy=-{\partial}_{y}{A_{x}(y,t)}dx{\wedge}dy=
2A^{0}[-2y^{-3}+y^{-4}][sinh(Ky^{-2})+cosh(Ky^{-2})]e^{{\gamma}t}dx{\wedge}dy
\label{24}
\end{equation}
One might take special attention to the boundary conditions of this field, which is spatially bounded at $\infty$ and grows without bound at the origin of the Lobachevsky plane. Note that if one compare the uniform magnetic field strength above, one observes that solution (\ref{24}) is a linear combination of uniform strengths in the form
\begin{equation}
B=B_{z}dx{\wedge}dy=2A^{0}[-2y^{-1}+y^{-2}][sinh(Ky^{-2})+cosh(Ky^{-2})]e^{{\gamma}t}B_{0}
\label{25}
\end{equation}
Note that the minus sign inside the brackets allows us to see that there is a magnetic field reversal below and above the line
\begin{equation}
-2y^{-1}+y^{-2}={0}
\label{26}
\end{equation}
which is equivalent to the line $y_{0}=\frac{1}{2}$ , parallel to the x-axis, as can be seen in any simple plotting of $B_{z}$ versus y. In the next section one shall compute the topological entropy of this example of fast dynamo along with the other forcing dynamo in 2D hyperbolic space. Actually $B_{z}$ goes to zero as one goes ${\infty}$. Now let us show that a non-geodesic equation, with an external force $F(y)$ can be obtained from this dynamo as
\begin{equation}
\frac{d^{2}x^{A}}{ds^{2}}+{{\Gamma}^{A}}_{BC}\frac{dx^{B}}{ds}\frac{dx^{C}}{ds}=F(y)
\label{27}
\end{equation}
where (B,C=1,2). Let us now use the following Riemann-Christoffel ${{\Gamma}^{i}}_{jk}$ components as
\begin{equation}
{{\Gamma}^{1}}_{12}={{\Gamma}^{2}}_{22}=-\frac{1}{y}
\label{28}
\end{equation}
\begin{equation}
{{\Gamma}^{2}}_{11}=\frac{1}{y}
\label{29}
\end{equation}
into (\ref{26}), with $V^{x}$ vanishing, to obtain the following non-geodesic equation
\begin{equation}
\frac{dV^{y}}{ds}+{{\Gamma}^{2}}_{22}(V^{y})^{2}=F^{y}(y)
\label{30}
\end{equation}
where the equation for $V^{x}$ is trivial. Together with the $div\textbf{v}=0$ equation this yields
\begin{equation}
{V^{y}}=y^{2}
\label{31}
\end{equation}
whose corresponding external force is
\begin{equation}
F(y)=-y
\label{32}
\end{equation}
which is a restoring type force on the dynamo flow. Note that expression (\ref{31}) reflects the parabolic or negative constant curvature nature of the flow.
\section{Restoring forces in non-geodesic dynamo flow in hyperbolic manifold}
In this section one shall drop the relation $curl(curlB)=-{\lambda}^{2}B$ used in the last section, and favor the forced non-geodesic dynamo flow in Lobachevsky plane. From this point of view one is led to use the expression (\ref{17}), and to compute the Laplacian ${\Delta}={\nabla}^{2}$ operator applied to the magnetic vector field, in terms of the above Riemann-Christoffel connection. Let us then expand these expressions in terms of the equation by
\begin{equation}
{A^{x;k}}_{;k}=[g^{jk}({A^{x}}_{,k}+{{\Gamma}^{x}}_{lk}A^{l})]_{,j}+g^{jl}{{\Gamma}^{x}}_{
kj}({A^{k}}_{,l}+{{\Gamma}^{k}}_{lm}A^{m})+g^{kl}{{\Gamma}^{j}}_{kj}({A^{x}}_{,l}+
{{\Gamma}^{x}}_{lm}A^{m})\label{33}
\end{equation}
which from the Riemann-Christoffel symbols yields
\begin{equation}
({\gamma}-{\eta}){A^{x}}=[{\eta}y-V_{)}y^{-4}]{A^{x}}_{,y}+{\eta}y^{2}{A^{x}}_{,yy}\label{34}
\end{equation}
where to simplify matters and obtain a very simple equation one assumed that $A^{y}$ and $V^{x}$ vanish identically. The divergence-free solenoidal incompressible flow condition yields
\begin{equation}
{A^{x}}(x,y,t)=y^{2}{A_{0}}(x,t)\label{35}
\end{equation}
which upon substitution into the self-induction equation determines the function $A_{0}(x,t)$ as
\begin{equation}
{A^{x}}(x,y,t)=y^{2}e^{x({\gamma}y^{-1}-V_{0}y^{-2})+{\gamma}t}\label{36}
\end{equation}
where one has used the following flow profile
\begin{equation}
V=V^{x}\frac{{\partial}}{{\partial}x}=V^{0}\frac{{\partial}}{{\partial}x}\label{37}
\end{equation}
Finally the magnetic field closed 2-form is given by
\begin{equation}
B=B_{z}dx{\wedge}dy=2y^{-2}xA_{x}({\gamma}-\frac{1}{2}V_{0}y^{-1})dx{\wedge}dy
\label{38}
\end{equation}
This magnetic field vanishes at the straight line ${\gamma}y_{0}=\frac{1}{2}V_{0}$ and since ${\gamma}$ has to be negative, otherwise the dynamo would be fast, the region $y_{0}=constant\ge{0}$ would gives us a negative $V_{0}$ and vice-versa. This shows that $y_{0}=0$ is an strange attractor for the magnetic flow and the field is also reversed. Note as well that when $y_{0}\ge{V_{0}}{2{\gamma}}$ the magnetic field is positive, otherwise it is negative inverting its sign as in geodynamos.
\section{Fast dynamo topological entropies in Lobachevsky plane}
Topological entropy is a fundamental ingredient in the construction of fast dynamo action \cite{10}. In this section one shall review the subject before stating and prove the main theorem of the paper. Two important theorems on dissipative and non-dissipative dynamos can be stated as \cite{10}:
\newline
\textbf{Theorem 1}:\newline
(Kozlov \cite{11}, Klapper and Young \cite{10}) Let v be a $div(v)=0$ $C^{\infty}$ vector field on a compact 3D manifold $\cal{M}$, and ${\mu}$ the Riemannian volume form on $\cal{M}$. Assume that a magnetic field $B_{0}$ is transported by the flow ${g_{v}}^{t}$ of the field v: $B(t)={g^{t}}_{v}*B_{0}$. Then for every continuos field $B_{0}$ on $\cal{M}$, the increment of the fast dynamo growth rate ${\gamma}$ has an upper bound given by the topological entropy $h_{top}(v)$ of the field v:
\begin{equation}
lim_{t\rightarrow{\infty}}sup\frac{1}{t}ln\int{||B(t)||{\mu}}\le{h_{top}(v)}
\label{39}
\end{equation}
This theorem for non-dissipative dynamos and mainly the next one on dissipative dynamos shall be used below to prove theorem 3.
\textbf{Theorem 2}:\newline
Let us be given a hyperbolic metric $g \in{\cal{M}}$ on a Riemannian manifold $\cal{M}$, and a closed magnetic two-form B as $B=dA$ (dB=0) where A is a magnetic vector potential one-form. If g is a Lobachevski Riemannian conformally flat metric whose line element is
\begin{equation}
ds^{2}=y^{-2}[dx^{2}+dy^{2}]\label{40}
\end{equation}
Thus a force-free non-geodesic dissipative dynamo flow associated with that magnetic flow possesses a positive topological entropy and topological entropy bounds non-geodesic dissipative dynamos in hyperbolic compact manifolds of negative constant Riemannian curvature.\newline
\textbf{Proof}: By assuming the last theorem and applying it to the force-free solution of the force-free fast dynamos of the last section yields
\begin{equation}
h_{top}\ge{{\gamma}(\eta)}={V}_{0}K-{\eta}{\lambda}^{2}\ge{0}
\label{41}
\end{equation}
since the last inequality implies the fast dynamo action, this implies that the component $V_{0}$ be bound by the relation, $V_{0}\ge{\frac{{\eta}{\lambda}^{2}}{K}}$. Note also that when the topological entropy vanishes. The computation of the magnetic energy
where ${\mu}$ is the Riemannian volume element, is given by
\begin{equation}
E(B)=\int{B^{2}{\mu}}\label{42}
\end{equation}
which by approximating the integral for $y\rightarrow{0}$, yields
\begin{equation}
E(B)=\int{B^{2}{\mu}}\approx{-\frac{1}{3}xy^{-3}ze^{{\gamma}t}}\label{43}
\end{equation}
which shows that the magnetic energy also grows spatially.
\section{Conclusions}
Geodynamo studies for the explanation of the magnetic field reversals of the magnetic field of the Earth has been one of the most important challenges in physics and mathematics. In this paper one makes use of a restoring force model of kinematic fast dynamo model in 2D manifolds of negative constant Riemannian curvature to explain magnetic field reversals. To easy the computations one makes use of the Lobachevsky hyperbolic plane. Besides the importance of the subject to physics and geophysics, from the mathematical side one also computes the topological entropy of the dynamo, so important for dynamical system theory. Turbulent dynamos in hyperbolic spaces, can be derived from this simple toy model and computer simulations by the PENCIL CODE \cite{12}.
\section{Acknowledgements}
Several discussions with A Brandenburg, D Sokoloff and Yu Latushkin are highly appreciated. I also thank financial  supports from UERJ and CNPq. Parts of this work started when the author was on two leaves of absence in Princeton university and in astrophysical dynamos program solar and stellar cycles and dynamos at NORDITA, Stockholm. Financial supports from both institutions are greatful acknowledged.
 

\newpage

\begin{thebibliography}{12}
  \bibitem{1} C. Chicone and Yu Latushkin, Evolution Semigroups in Dynamical systems and differential equations, American
  Mathematical Society, AMS-(1999). C. Chicone and Yu Latushkin and S. Montgomery-Smith, Comm. Math. Physics \textbf{173} 379 (1995). C. Chicone and Yu Latushkin, Proc of the American Mathematical Society 125, N. 11,3391 (1997). 
  \bibitem{2} V. Arnold, Ya B. Zeldovich, A. Ruzmaikin and D.D.
  Sokoloff, JETP 81 (1981),n. 6, 2052. V. Arnold, Ya B. Zeldovich, A. Ruzmaikin and D.D.
  Sokoloff, Doklady Akad. Nauka SSSR 266 (1982) n6, 1357.
  \bibitem{3} Petrelis, Phys Rev \textbf{E 80},035302 (2009).
  \bibitem{4} A Shukurov, R Stepanov, D Sokoloff, Phys Rev E (2008).
  \bibitem{5} R Klebanov, J Maldacena, Phys Today (2008).
  \bibitem{6} S Carneiro, Gen Rel Grav J (2002).
  \bibitem{7} L C Garcia de Andrade, Lotentz tori AdS fast dynamos in hyperbolic spaces, (2009) arxiv preprint.
  \bibitem{8} T Sunada, Magnetic Flows on a Riemann Surface, Proc Kaist Math Workshop (1993). 
  \bibitem{9} C Tsagas and R Maartens, Class and Quantum Gravity \textbf{17},(2002),2215.
  \bibitem{10} I Klapper and L Young, Comm Math Phys \textbf{173}(1995),623.
  \bibitem{11} V Kozlovsky, Ergodic theory and Dynamical systems,(1997). 
  \bibitem{12} A Brandenburg and Garcia de Andrade, Turbulent dynamos in hyperbolic spaces, in preparation.
    \end{thebibliography}
  \end{document}